\documentclass[a4paper, 11pt,english]{article}

\usepackage[T1]{fontenc}

\usepackage{amsmath}
\usepackage{amstext}
\usepackage{wasysym}
\usepackage{esint}
\usepackage{babel}
\usepackage{lscape}
\usepackage{cancel}
\usepackage{bm}
\usepackage{bbm}
\usepackage{multirow} 
\usepackage{amsfonts}
\usepackage{amssymb}
\usepackage{graphicx}
\usepackage[usenames,dvipsnames]{color}
\usepackage{hyperref}
\usepackage{epstopdf,epsfig}
\usepackage{verbatim}
\usepackage[utf8]{inputenc}
\usepackage[left=1in,right=1in,top=1in,bottom=1in]{geometry}             % For good copy.
\usepackage{booktabs} % for much better looking tables
\usepackage{array} % for better arrays (eg matrices) in maths
\usepackage{paralist} % very flexible & customisable lists (eg. enumerate/itemize, etc.)
\usepackage{verbatim} % adds environment for commenting out blocks of text & for better verbatim
\usepackage{subfig} % make it possible to include more than one captioned figure/table in a single float
\hypersetup{
    colorlinks=true,         % false: boxed links; true: colored links, false is default
    linkcolor=blue,          % color of internal links, red is default
    citecolor=red,        % color of links to bibliography, 'green' is default
    urlcolor=Violet             % color of external links, cyan is default
}

\begin{document}
\title{Background independent holographic dual to $T\bar{T}$ deformed CFT with large central charge in 2 dimensions}
\author{{\bf Vasudev Shyam}\footnote{\href{mailto:vshyam@pitp.ca}{vshyam@pitp.ca}} \\\it Perimeter Institute for Theoretical Physics\\ \it 31 N. Caroline St. Waterloo, ON, N2L 2Y5, Canada }
\date{\today}
\maketitle

\begin{abstract}
The geometerization of the renormalization group flow triggered by the $T\bar{T}$ deformation of large $c$ conformal field theories in two dimensions is presented. This entails the construction of the off shell Einstein-Hilbert action in three dimensions from said renormalization group flow. 

The crucial ingredient to this construction will be the encoding of general covariance in the emergent bulk theory in a very particular form of the Wess--Zumino consistency conditions. The utilisation of the local renormalization group, which requires putting the theory under consideration on an arbitrary background geometry, supplemented by the aforementioned covariance condition ensures that the whole construction is background independent.

\end{abstract}

\begin{section}{Introduction}
One of the important questions for the theory of quantum gravity to answer is the mechanism by which space and time emerge from the dynamics of the fundamental theory operating at the Planck scale. Holography provides a partial answer through the emergence of a direction of space starting from a quantum field theory living in one lower dimension. Constructing the bulk theory dual to a given `holographic' quantum field theory demonstrates how such a dimension of space emerges concretely. This article will focus on such a construction of the holographic dual to a particular kind integrable deformation of conformal field theories in two dimensions with large central charge. This deformation consists a product of the holomorphic and anti holomorphic components of the energy momentum tensor, and is thus irrelevant. 

The geometerization of the renormalization group is one of the defining features of the holographic duality. The duality itself can be seen as a mapping of the renormalization group flow of a quantum field theory (in some limit) into the classical equations of motion of a theory of gravity coupled to other fields living in one higher dimension than the field theory. The field theory is thought of to live on the boundary of the gravity theory living in the `bulk'. The renormalization group scale is identified with the extra dimension of the bulk theory. This way the extra bulk dimension is said to be emergent from the boundary quantum field theory's RG flow. These ideas are the backbone of what is known as the holographic renormalization group applied to theories known to possess local holographic duals (see for instance \cite{HRG} or \cite{Papa2} for a more recent overview).

In this work, such a geometerization of the RG flow triggered by the so called $T\bar{T}$ operator in two dimensional conformal field theories with large central charge is presented. This will require the renormalization group flow to be cast into the dynamics of general relativity in one higher dimension. This happens through constructing the off shell action for the bulk gravity theory. The tool used for this mapping is known as quantum RG (\cite{S.S.Lee1},\cite{S.S.Lee2}). This however need be supplemented with additional consistency conditions. These conditions stem from the demand that the bulk theory be generally covariant.  In order for this to be the case the emergent bulk direction must be treated on equal footing with those of the boundary theory.  

I will provide a rough description of the techniques utilised for this quantum RG mapping to transform renormalization group flows into a dynamical theory in one higher dimension. 
We are typically interested in field theories in the bulk, so in order for the renormalization group flow equations to mimic equations of motion, coupling constants must be upgraded to sources which depend on space. The coarse graining transformations in all generality can act through Weyl transformations of the background metric. This is much akin to how coarse graining is manifested through Kadanoff's blocking transformation on lattice systems. The response of the generating functional of the quantum field theory on said background to such a Weyl transformation shall determine the flow equations and is thus the local variant of the Callan--Symanzik equation. This approach is known as the local renormalization group, due to Osborn et. al. \cite{Osborn1}. 

This local renormalization group is typically supplemented with consistency conditions that encode the commutativity of the local Weyl transformations of the background metric. 
These consistency conditions are known as the Wess--Zumino conditions:
$$[\Delta_{\sigma},\Delta_{\sigma'}]\textrm{ln}Z=0,$$
where $Z$ denotes the partition function of the theory of interest and $\Delta_{\sigma}$ is the generator of infinitesimal local RG transformations. 
The property of general covariance in the bulk theory is encoded as a particular form of the Wess--Zumino consistency conditions. As I will argue in section 3, theories possessing local bulk duals necessarily satisfy this condition- which I call the holographic Wess--Zumino consistency condition \cite{me}. This condition will be linchpin to properly constructing the bulk gravity theory off shell in the large central charge limit of the conformal field theory deformed by a composite operator of the stress tensor mentioned earlier. 

The article is organised as follows. Section 2 is devoted to laying out the field theoretic starting point upon which the rest of the work builds. This includes two dimensional conformal field theories on curved backgrounds and a preliminary description of the $T\bar{T}$ deformation and some of its properties such as the factorisation of its expectation value. Concepts relating to local renormalization group pertinent to the construction of the bulk theory, such as reinterpreting a certain anomalous Ward identity as the local Callan--Symanzik equation are also described. The construction of the bulk gravity theory through the quantum RG mapping of the RG flow of the $T\bar{T}$ deformed large $c$ CFT is described in section 3. The holographic Wess--Zumino consistency condition and how it relates to the Poisson algebra of constraints of the bulk theory is described and the role it plays in constraining the form of the bulk Hamiltonian is emphasised. The Hamilton--Jacobi equations are derived and the manner in which the cosmological constant arises in the bulk through a canonical transformation of the bulk Hamiltonian is presented. Before concluding the introduction, I will provide comparison of what is done in the article to previous related work.

\subsection{Comparison to earlier work}
The holographic interpretation of the $T\bar{T}$ deformation was first presented in the work of  McGough, Mezei and Verlinde \cite{Ver} and among other non trivial checks, the dictionary between an exact RG equation and the Einstein Hamilton Jacobi equations in 3 dimensions was conjectured. The work presented here is a construction of the off shell bulk theory that substantiates the aforementioned conjecture, and in a sense `derives' the bulk Hamilton Jacobi Equations. The method used to accomplish this, known as the quantum renormalization group (QRG) due to Sung-Sik Lee, was applied to similar effect in the context of large N matrix field theories in a hypothetical limit where the single trace energy momentum tensor is the only operator with finite scaling dimension in \cite{S.S.Lee1}, \cite{S.S.Lee2}. The general methods however which involve manipulations of functional integrals can be applied more widely as is done in this work. The key ingredient that the aforementioned works on QRG were missing which provides a guarantee that the bulk theory in the semiclassical or large $N$ limit indeed describes Einstein gravity is the mechanism by which general covariance ought to emerge. This is codified in the Holographic Wess--Zumino consistency conditions introduced in \cite{me}. The following work can be seen as an application of this idea to large $c$ CFTs in two dimensions deformed by an irrelevant operator. 
\end{section}

\section{The $T\bar{T}$ deformed conformal field theories}
Before delving into the holographic construction of the three dimensional bulk theory, I will first describe the field theory of interest, namely the quantum field theory resulting from the deformation of a large $c$ CFT in two dimensions by the operator $T\bar{T}$ (whose form I will later specify). I begin by laying out the conventions and notation pertaining to two dimensional conformal field theories on arbitrary curved backgrounds. 
\subsection{2D conformal field theories on curved backgrounds}
The conformal field theories under consideration are those living in two dimensional spacetime. For my purposes I will work with Euclidean conformal field theories. I will denote the action of the conformal field theory as $S_{CFT}$ and the partition function reads
\begin{equation}
Z_{CFT}[g]=\int \mathcal{D}\Phi e^{-S_{CFT}[\Phi;g]}.
\end{equation}
Here $g_{\mu\nu}$ denotes the metric tensor on the two dimensional space on which the conformal field theory lives. The $\Phi$s refer to fundamental fields of arbitrary spin in principle. They needn't be specific for the rest of the discussion.

 Putting the theory on an arbitrary curved background with metric $g_{\mu\nu}$ means that conformal symmetry will be anomalous. The anomaly itself is the sole source of scale dependence in this theory, and hence it will be useful to isolate said scale dependence. It will be useful to note that the metric can be decomposed as follows:
$$g_{\mu\nu}=e^{2\varphi(x)}\hat{g}_{\mu\nu},$$
where $\varphi(x)$ is referred to as the Liouville field. 
The partition function reflects this decomposition in the following manner:
\begin{equation}Z_{CFT}[g=e^{2\varphi}\hat{g}]=e^{S_{P}[\varphi;\hat{g}]}Z^{*}_{CFT}[\hat{g}].\end{equation}
The scale dependence rests in the Polyakov action 
$$S_{P}[\varphi;\hat{g}]=\int \textrm{d}^{2}x \sqrt{\hat{g}}\left(\varphi \square \varphi-\varphi \bar{R}(\hat{g})\right).$$
The Ricci scalar $R$ for $g$ and that of $\hat{g}$ (denoted $\bar{R}(\hat{g})$) are related in the following manner:
\begin{equation*}R(g=\hat{g}e^{2\varphi(x)})=e^{2\varphi}(\bar{R}(\hat{g})+2\square\varphi)\end{equation*}
The defining property of the conformal field theory is the Ward identity corresponding to Weyl invariance which reads:
\begin{equation}\left(\frac{\delta}{\delta \varphi}-\frac{c}{24\pi}e^{-2\varphi}R(g)\right)Z_{CFT}[g]=0.\end{equation}
Here, $c$ denotes the central charge. 

The purpose of isolating the $\varphi$ dependence is to highlight the fact that the Polyakov action is solely responsible for the scale anomaly:
\begin{equation}\frac{\delta S_{p}[\varphi;\hat{g}]}{\delta \varphi}=-\frac{c}{24\pi}e^{-2\varphi}R(g),\end{equation} 
where as $Z^{*}_{CFT}[\hat{g}]$ remains completely Weyl invariant. 

Despite the Weyl anomaly, the aforementioned ward identity will still be referred to as the conformal Ward identity. The breaking of conformal symmetry which results from RG flows being triggered will alter the form of this Ward identity, and the form the identity then takes will define an exact RG flow equation. 

\subsection{The deforming operator $T\bar{T}$}
The form of the deforming operator $\mathcal{O}$ (which I referred to previously as $T\bar{T}$, and shall continue to use these two notations interchangeably) on flat space with complex co-ordinates $(z,\bar{z})$ is
$$\mathcal{O}=T\bar{T}-\frac{1}{4}\Theta^{2},$$
where $T,\bar{T},\Theta$ stand for the holomorphic, antiholomorphic components and trace of the energy momentum tensor of the theory. In the limit where the theory is indeed conformal, $\mathcal{O}=T\bar{T}$.
 Zamalodchikov in \cite{Zamo1},\cite{Zamo2} proved that this deformation is integrable. This means that it is one of an infinite number of mutually commuting conserved charges. It was also shown that in the flat space limit for a wide class of slowly varying, translation invariant states, the following factorisation property holds for the expectation value of this operator:
\begin{equation}
\langle \mathcal{O}\rangle=\langle T\rangle \langle \bar{T}\rangle -\langle \Theta \rangle^{2}. \label{fact}
\end{equation}
This will be useful in constructing the bulk dual to this theory. Clearly, this operator is irrelevant, and as the above factorisation property suggests, it is of dimension four. Thus the coupling of this operator in the action of the perturbed field theory is of mass dimension -2.
It was shown in \cite{Dubov} that deforming a two dimensional quantum field theory by this operator on flat space is equivalent to coupling the theory to Jackiw--Teitelboim gravity. This is a very interesting observation but for the purposes of this article, i.e. to construct the off shell Einstein-Hilbert action in the holographically dual bulk theory, it will be necessary to put the theory we are interested in on an arbitrary curved background. In that case, $\mathcal{O}$ in a more covariant form is given by
\begin{equation}\mathcal{O}=-\frac{1}{8}\left(T_{\mu\nu}T^{\mu\nu}-(T^{\alpha}_{\alpha})^{2}\right),\end{equation}
it can also be rewritten by noting that 
$$T_{\mu\nu}T^{\mu\nu}-(T^{\alpha}_{\alpha})^{2}=G^{\textrm{dW}}_{\mu\nu\rho\sigma}T^{\mu\nu}T^{\rho\sigma},$$
where the de Witt super metric is defined as
\begin{equation}G^{\textrm{dW}}_{\mu\nu\rho\sigma}=(g_{\mu(\rho}g_{\sigma)\nu}-g_{\mu\nu}g_{\rho\sigma}).\end{equation}

The desired factorisation property \eqref{fact} will now take the form

\begin{equation*}\langle G^{\textrm{dW}}_{\mu\nu\rho\sigma}T^{\mu\nu}T^{\rho\sigma} \rangle = G^{\textrm{dW}}_{\mu\nu\rho\sigma}\langle T^{\mu\nu}\rangle \langle T^{\rho\sigma}\rangle.\end{equation*}
And in order for the dual theory to indeed be general relativity in three dimensions, we will see that the above property has to be exact in addition to the fact that no operator other than $\mathcal{O}$ is generated to leading order in the coupling in front of it in conformal perturbation theory. In the next section, I will show that there is a natural consistency condition that the quantum field theory must satisfy in order for these hopes to be realised. 

Now to take a look at the local Callan--Symanzik equation. 
\subsection{The local renormalization group equation}

The theory of interest in the discussion to follow has the following action
\begin{equation}S[\Phi;g,\mu]=S_{CFT}[\Phi;g]+\frac{\mu}{16}\int \textrm{d}^{2}x \sqrt{g} G^{\textrm{dW}}_{\mu\nu\alpha\beta}T^{\mu\nu}T^{\alpha\beta},\end{equation}
where $\mu$ is kept infinitesimal for the rest of the discussion. This theory is no longer conformal, but depends on an additional scale through $\mu$. The geometerization of the renormalization group flow this operator generates will entail said flow being mapped to Einstein's equation in a three dimensional (Euclidean) spacetime.  

The aim here is to write down a Callan--Symanzik like equation describing the response of the generating functional to local Weyl transformations. These Weyl transformations are to be seen as a continuum generalisation of the blocking transformations of Kadanoff, so they encode coarse graining but in a space dependent manner. 

To begin, the partition function of the deformed field theory takes the form
\begin{equation}
Z_{QFT}[g,\mu]=Z_{CFT}[g]\langle e^{-\frac{\mu}{2}\int \textrm{d}^{2}x\sqrt{g} G^{\textrm{dW}}_{\mu\nu\rho\sigma}T^{\mu\nu}T^{\rho\sigma}}\rangle_{_{CFT} }.
\end{equation}
The key results of this subsection shall all follow from the above expression where $\mu$ is considered to be small. It will thus be very important to state the assumptions that go behind this expression. 

The most important one is that the only scale in the quantum field theory is $\mu$. This implicitly assumes that no operators of higher dimensions suppressed by other scales enter into the RG flow of the quantum field theory. I will assume that the theory possesses large central charge $c$. Given that the energy momentum tensor and its derivatives, i.e. the descendants of the identity operator close under the operator product expansion with each other, the higher dimension operators possibly entering with scales other than $\mu$ are all composite operators of the energy momentum tensor. In the next section, I will describe what the necessary condition is in order to protect the theory of interest from such corrections while retaining only $\mathcal{O}$. For the time being, I will continue with more the more stringent assumption that $\mu$ is in fact the only scale available in the quantum field theory, and also that $g_{\mu\nu}(x)$ to vary slowly with $x$. 

I will start by introducing the Liouville field: 
\begin{equation*}
Z_{QFT}[g,\mu]=Z^{*}_{CFT}[\hat{g}]e^{S_{P}[\hat{g},\varphi]}\langle e^{-\frac{\mu}{2}\int \textrm{d}^{2}x\sqrt{\hat{g}}e^{-2\varphi} G^{\textrm{dW}}_{\mu\nu\rho\sigma}T^{\mu\nu}T^{\rho\sigma}}\rangle_{_{CFT} }.
\end{equation*}
The modified scale Ward identity that takes into account the breaking of conformal invariance by the deformation-
$$-\frac{e^{-2\varphi}}{4}\frac{\delta \textrm{ln}Z_{QFT}[g,\mu]}{\delta \varphi}=\langle T^{\alpha}_{\alpha}\rangle =-\frac{c}{96\pi}R(\varphi,\hat{g})-\frac{\mu}{2}\langle G^{\textrm{dW}}_{\mu\nu\rho\sigma}(\varphi,\hat{g})T^{\mu\nu}T^{\rho \sigma}\rangle,$$
which shows how in addition to the anomaly, how the operator $\mathcal{O}$ drive the RG flow, at least to the lowest order in $\mu$.
 Another use for the Liouville field is to act as a compensator for Weyl transformations- where transformations of the form $g_{\alpha \beta}\rightarrow e^{2\sigma}g_{\alpha\beta}$, this can be compensated through $\varphi\rightarrow \varphi-\sigma$. Thus the above equation can be written as 
\begin{equation}\delta_{\sigma}\textrm{ln}Z=\int \textrm{d}^{2}x \sqrt{g}\sigma \langle T^{\alpha}_{\alpha} \rangle= \int \textrm{d}^{2}x\sqrt{g}\sigma \left(-\frac{c}{96\pi}R(g)-\frac{\mu}{2}\langle G^{\textrm{dW}}_{\mu\nu\rho\sigma}T^{\mu\nu}T^{\rho \sigma}\rangle\right),\label{lcsb}\end{equation}
where $\delta_{\sigma}:=\int\textrm{d}^{2}x \sqrt{g}\sigma g_{\mu\nu}\frac{\delta}{\delta g_{\mu\nu}}.$
 Notice that despite the split between the Liouville field and the metric $\hat{g}_{\mu\nu}$ the right hand side of the above equation only sees $\varphi$ and $\hat{g}_{\mu\nu}$ in their combination $g_{\mu\nu}$. This is an equation which describes the response of the generating functional to the change of local scale encoded in the Weyl transformation $g_{\mu\nu}\rightarrow e^{-2\sigma(x)}g_{\mu\nu}$. This Weyl transformation can be seen as a generalisation of the blocking transformation Kadanoff introduced on the lattice. In principle, if other sources were considered, then functional derivatives with respect to such sources will result in correlation functions of the operators to which they couple. The response of such correlation functions under the aforementioned coarse graining transformations corresponds to inserting the trace of the energy momentum tensor into said correlates which one can deduce from the above equation. Thus this equation is the local Callan--Symanzik equation. 

Now, in the large $c$ limit and taking  $\hat{g}_{\mu\nu}=\eta_{\mu\nu}$, the factorisation property of Zamalodchikov can be utilised:
\begin{equation}
\langle \Theta \rangle=  -\frac{c}{96\pi}R(\varphi)-\frac{\mu}{2}\left(\eta_{\mu(\rho}\eta_{\sigma)\nu}-\eta_{\mu\nu}\eta_{\rho\sigma}\right) \langle T^{\mu\nu}\rangle\langle T^{\rho\sigma}\rangle. \label{lcs}
\end{equation}
Notice that $\frac{1}{2}\left(\eta_{\mu(\rho}\eta_{\sigma)\nu}-\eta_{\mu\nu}\eta_{\rho\sigma}\right)\langle T^{\mu\nu}\rangle\langle T^{\rho\sigma}\rangle=\langle T \rangle\langle \bar{T} \rangle-\langle \Theta \rangle^{2}$ in the $(z,\bar{z})$ co-ordinates. 

The local RG equation \eqref{lcs} can be seen as the defining property that the quantum field theory perturbed by $\mathcal{O}$ needs to satisfy. But we must remember that several assumptions were made to derive this equation, and the aim of the next section will be to present a consistency condition which will allow one to relax a few of these assumptions and also to allow the construction of the Einstein-Hilbert action of the holographically dual bulk theory. In other words, this consistency condition too must be included in the very definition of the theory we are interested in. 

 The last subsection to follow will describe why \eqref{lcs} would ensure that it is indeed general relativity operating in the bulk theory.  
 \subsection{What to aim at: The holographic interpretation of the local Callan-Zymanzik equation} 
 If the field theory under consideration were to possess a holographic dual, then the generating function can be equated, at large $c$ to the on shell classical Einstein--Hilbert action for gravity in $2+1$ dimensions (denoted $S_{c}$),
\begin{equation}\textrm{ln}Z_{QFT}[g;\mu]=-\frac{c}{24\pi}S_{c}[g,\mu]\label{aim}\end{equation}
The local Callan--Symanzik equation for such a theory takes the form
\begin{equation}
\frac{\delta S_{c}}{\delta \varphi}=e^{-2\varphi} R(\varphi,\hat{g})-\frac{c\mu}{24\pi}\left(\frac{\delta S_{c}}{\delta \hat{g}_{\mu\nu}}\frac{\delta S_{c}}{\delta \hat{g}_{\rho\sigma}}-\frac{1}{2}\left(\frac{\delta S_{c}}{\delta \varphi}\right)^{2}\right).
\end{equation} 
This is nothing but the Einstein-Hamilton-Jacobi equation for pure (Euclidean) general relativity in $2+1$ dimensions provided one sets $\mu=\frac{24\pi}{c}$. This observation was made in \cite{Ver}, where the above expression first appeared. Notice however that in contrast to \eqref{lcs}, the above equation holds away from $\hat{g}_{\mu\nu}=\eta_{\mu\nu}$, and beyond the leading order in the gradient expansion. It is instructive to rewrite the above equation in terms of the metric $g_{\mu\nu}$:
\begin{equation}
g_{\mu\nu}\frac{\delta S_{c}}{\delta g_{\mu\nu}}=-R(g)-G^{\textrm{dW}}_{\mu\nu\rho\sigma}\frac{\delta S_{c}}{\delta g_{\mu\nu}}\frac{\delta S_{c}}{\delta g_{\rho\sigma}}.
\end{equation}
Recalling that $\frac{\delta S_{c}}{\delta g_{\mu\nu}}=\langle T^{\mu\nu}\rangle$, and comparing the above form of the local Callan--Symanzik equation with the general form \eqref{lcsb}, we see that the factorisation condition
$$\langle G^{\textrm{dW}}_{\mu\nu\rho\sigma} T^{\mu\nu} T^{\rho\sigma}\rangle =G^{\textrm{dW}}_{\mu\nu\rho\sigma}\langle T^{\mu\nu}\rangle\langle T^{\rho\sigma}\rangle$$
holds for these theories that possess holographic duals. This is what I meant earlier when I mentioned that such a factorisation property will be crucial importance. 
 
The remainder of this article is dedicated to deriving the above result \eqref{aim} starting from the $T\bar{T}$ deformed boundary CFT by reorganising the RG flow. Such efforts to construct the bulk theory from the RG flow of the boundary theory go under the heading of the `quantum renormalization group' (QRG) \cite{S.S.Lee1}, \cite{S.S.Lee2}. An intended by product of such an effort will be to clearly understand what criteria the theories which possess holographic bulk duals should possess.

\section{Constructing the Bulk theory}
In this section, I wish to start only from the two dimensional field theory deformed by the operator $\mathcal{O}$ and construct from that the off shell, Einstein Hilbert action in three dimensions. I will show that the criterion for such a construction to work will hinge on a very particular composition property of the coarse graining transformations. 

In other words, I wish to show what criterion need be met in order for the $\mathcal{O}$ deformed quantum field theory to remain protected under corrections involving operators of higher dimensions including gradients etc. at least in the large $c$ limit and in conformal perturbation theory. In other words, many of the assumptions from the previous section can be dropped provided that this criterion is met. 

The idea behind the holographic duality is that the evolving bulk fields (are a subset of) the sources of the boundary operators somehow granted dynamics. More specifically, the radial evolution of the bulk fields is equated with the renormalization group flow. In the case of interest here, the conformal invariance is broken by a composite operator involving solely the energy momentum tensor, whose source is the metric. On the other side of the duality, the only evolving field will thus be the metric, and so one can expect pure gravity to be the theory at play in the bulk. In the first subsection I describe how to grant dynamics to sources of composite operators in the theory. In the case of interest, we have just the metric and that couples to the energy momentum tensor.

\begin{subsection}{The dual theory}
The end result of the QRG procedure applied to the theory we are interested in, which is a two dimensional conformal field theory with large central charge $c$ deformed by the $T\bar{T}$ operator. The path integral manifestation of the anomalous Ward identity is that it allows us to answer the question of how the functional integral responds to the infinitesimal Weyl transformations $g_{\mu\nu}\rightarrow e^{\sigma^{(0)}(x) \delta z}g_{\mu\nu}$, where $\delta z$ is an infinitesimal constant. This is given by
\begin{equation}
Z[g_{\mu\nu}]=\int \mathcal{D}\Phi e^{-S_{QFT}[\Phi]-\delta z\int_{x}\sigma^{(0)}(x)(T^{\alpha}_{\alpha}-\frac{\mu}{16}G^{\textrm{dW}}_{\mu\nu\rho\eta}T^{\mu\nu}T^{\rho\eta}-\frac{c}{24\pi}R+\cdots)},
\end{equation}
and the ($\cdots$) stands for the potential corrections coming from potential higher dimension operators that might arise. This was the possibility alluded to in the previous section. In order to show how the consistency conditions I will later impose protect the form of the local Callan--Symanzik equation that can be interpreted as the Einstein Hamilton Jacobi equations, I will begin by assuming that higher order corrections too can in principle appear but all but the lowest order term corresponding to $\mathcal{O}$ will not satisfy the aforementioned criterion. To begin however, let $f(T^{\mu\nu},g_{\mu\nu})$ encode all these corrections i.e.
$$f(T^{\mu\nu},g_{\mu\nu}):=\frac{\mu}{16}G^{\textrm{dW}}_{\mu\nu\rho\eta}T^{\mu\nu}T^{\rho\eta}-\frac{c}{24\pi}R+\cdots$$
Now in order to grant the metric dynamics, we use the following trick:
\begin{equation*}
\int \mathcal{D}\Phi \mathcal{D}g^{(0)}_{\mu\nu} \delta(g^{(0)}_{\mu\nu}-g_{\mu\nu})   e^{-S_{QFT}[\Phi;g]-\delta z\int_{x}\sigma^{(0)}(x)(T^{\alpha}_{\alpha}-f(T^{\mu\nu},\ g^{(0)}_{\mu\nu}))}.
\end{equation*}
It will also be useful to write think of $T^{\mu\nu}$ as a normalised functional derivative with respect to the metric acting on the generating functional:
\begin{equation*}
\int \mathcal{D}\Phi \mathcal{D}g^{(0)}_{\mu\nu} \delta(g^{(0)}_{\mu\nu}-g_{\mu\nu})  \exp\left\{-\delta z\int_{x}\sigma^{(0)}(x)\left(\frac{24\pi}{c} g^{(0)}_{\mu\nu}\frac{\delta}{\delta g^{(0)}_{\mu\nu}}-f\left(\frac{24\pi}{c}\frac{\delta}{\delta g^{(0)}_{\mu\nu}},g^{(0)}_{\mu\nu})\right)\right)\right\}e^{-S_{QFT}[\Phi;g^{(0)}]}.
\end{equation*}

and then we exponentiate the delta function 
\begin{equation*}
 \int \mathcal{D}\Phi \mathcal{D}g^{(0)}_{\mu\nu} \mathcal{D}\pi^{(0)\mu\nu}  e^{-\frac{c}{24\pi}\pi^{(0)\mu\nu}(g^{(0)}_{\mu\nu}-g_{\mu\nu}) }  \times \end{equation*} \begin{equation*} \exp\left\{-\delta z\int_{x}\sigma^{(0)}(x)\left(\frac{24\pi}{c}g^{(0)}_{\mu\nu}\frac{\delta}{\delta g^{(0)}_{\mu\nu}}-f\left(\frac{24\pi}{c}\frac{\delta}{\delta g^{(0)}_{\mu\nu}},g^{(0)}_{\mu\nu}\right)\right)\right\}e^{-S_{QFT}[\Phi;g^{(0)}]}.
\end{equation*}
At this stage, we can also make note of the fact that the theory coupled to an arbitrary background geometry $g^{(0)}_{\mu\nu}$ should be invariant under diffeomorphism transformations $g^{(0)}_{\mu\nu}\rightarrow g^{(0)}_{\mu\nu}+ \delta z(\nabla_{(\mu}\xi^{(0)}_{\nu)})$. This invariance manifests itself through the diffeomorphism Ward identity which in the path integral appears as:
\begin{equation*}
 \int \mathcal{D}\Phi \mathcal{D}g^{(0)}_{\mu\nu} \mathcal{D}\pi^{(0)\mu\nu}  e^{-\frac{c}{24\pi}\pi^{(0)\mu\nu}(g^{(0)}_{\mu\nu}-g_{\mu\nu}) }  \times \end{equation*} \begin{equation*} \exp\left\{-\delta z\int_{x}\sigma^{(0)}(x)\left(\frac{24\pi}{c}g^{(0)}_{\mu\nu}\frac{\delta}{\delta g^{(0)}_{\mu\nu}}-f\left(\frac{24\pi}{c}\frac{\delta}{\delta g^{(0)}_{\mu\nu}},g^{(0)}_{\mu\nu}\right)\right)+\frac{24\pi}{c}(\nabla_{(\mu}\xi^{(0)}_{\nu)})\frac{\delta}{\delta g^{(0)}_{\mu\nu}}\right\}e^{-S_{QFT}[\Phi;g^{(0)}]}.
\end{equation*}
We can then functionally integrate by parts so that the functional derivatives with respect to the metric $g^{(0)}_{\mu\nu}$ get replaced by $\pi^{(0)\mu\nu}$, and we find
\begin{equation*}
 \int \mathcal{D}\Phi \mathcal{D}g^{(0)}_{\mu\nu} \mathcal{D}\pi^{(0)\mu\nu}  e^{-\frac{c}{24\pi}\pi^{(0)\mu\nu}(g^{(0)}_{\mu\nu}-g_{\mu\nu}) } \times \end{equation*} \begin{equation*}   \exp\left\{-\delta z\left(\frac{c}{24\pi}\right)\int_{x}\sigma^{(0)}(x)\left(g^{(0)}_{\mu\nu}\pi^{(0)\mu\nu}-f\left(\pi^{(0)\mu\nu},g^{(0)}_{\mu\nu}\right)\right)+(\nabla_{(\mu}\xi^{(0)}_{\nu)})\pi^{(0)\mu\nu}\right\}e^{-S_{QFT}[\Phi;g^{(0)}]}.
\end{equation*}
We see that the partition function of the original quantum field theory can be recollected although the theory now lives on the geometry described by metric $g^{(0)}_{\mu\nu}$:
\begin{equation*}
 \int \mathcal{D}g^{(0)}_{\mu\nu} \mathcal{D}\pi^{(0)\mu\nu}  e^{-\frac{c}{24\pi}\pi^{(0)\mu\nu}(g^{(0)}_{\mu\nu}-g_{\mu\nu}) } \times \end{equation*} \begin{equation*}   \exp\left\{-\delta z\left(\frac{c}{24\pi}\right)\int_{x}\sigma^{(0)}(x)\left(g^{(0)}_{\mu\nu}\pi^{(0)\mu\nu}-f\left(\pi^{(0)\mu\nu},g^{(0)}_{\mu\nu}\right)\right)+(\nabla_{(\mu}\xi^{(0)}_{\nu)})\pi^{(0)\mu\nu}\right\}Z_{QFT}[g^{(0)}].
\end{equation*}

Now this process of performing infinitesimal Weyl transformations can be iterated say $k$ times until we find
\begin{equation*}
 \int \prod^{k}_{i=0} \mathcal{D}g^{(i)}_{\mu\nu} \mathcal{D}\pi^{(i)\mu\nu}  e^{-\frac{c}{24\pi}k\delta z\sum^{k}_{i=0}\pi^{(i)\mu\nu}\frac{(g^{(i)}_{\mu\nu}-g^{(i-1)}_{\mu\nu})}{\delta z} } \times \end{equation*} \begin{equation*}   \exp\sum^{k}_{i=0}\left\{-\delta z\left(\frac{c}{24\pi}\right)\int_{x}\sigma^{(i)}(x)\left(g^{(i)}_{\mu\nu}\pi^{(i)\mu\nu}-f\left(\pi^{(i)\mu\nu},g^{(i)}_{\mu\nu}\right)\right)+(\nabla_{(\mu}\xi^{(i)}_{\nu)})\pi^{(i)\mu\nu}\right\}Z_{QFT}[g^{(k)}].
\end{equation*}

The continuum limit can now be taken where $\delta z\rightarrow 0$, $k\delta z \sum^{k}_{i=0}:=\int^{z_{*}} \textrm{d}z$, and the collections of fields $(g^{(i)}_{\mu\nu}(x),\pi^{(i)\mu\nu}(x),\sigma^{(i)}(x),\xi^{(i)\mu}(x))$ now are replaced by fields with $z$ dependence $(g_{\mu\nu}(x,z),\pi^{\mu\nu}(x,z),\sigma(x,z),\xi^{\mu}(x,z))$. Then we find the emergence of a `bulk theory' with action $S_{B}$:
\begin{equation*}
 \int \mathcal{D}g_{\mu\nu}(x,z) \mathcal{D}\pi^{\mu\nu}(x,z)\mathcal{D}\sigma(x,z)\mathcal{D}\xi_{\mu}(x,z)  e^{-\frac{c}{24\pi}S_{B}[g_{\mu\nu(x,z)},\pi^{\mu\nu}(x,z),\alpha(x,z),\xi^{\mu}(x,z)]}Z_{QFT}[g(x,z=z_{*})].
\end{equation*}
Here, the $\sigma(x,z)$ and $\xi(x,z)$ are also integrated over because they appear only linearly in the action and are thus  Lagrange multipliers. 

The form of the bulk action is given by
\begin{equation}
S_{B}=\int \textrm{d}^{2}x\textrm{d}z \sqrt{g}\left(\pi^{\mu\nu}(x,z)\dot{g}_{\mu\nu}(x,z)-H_{B}(\pi^{\mu\nu},g_{\mu\nu},\sigma,\xi^{\mu})\right).
\end{equation}
We see that the normalisation of the functional derivatives that get traded for the momenta and thus the overall normalisation of the action is chosen such that the large $c$ limit of this theory does indeed give the semiclassical limit where the partition function can be evaluated in the saddle point approximation. Now, the bulk Hamiltonian reads
$$H_{B}(\pi^{\mu\nu},g_{\mu\nu},\sigma,\xi^{\mu})=\sigma(x,z)H(\pi^{\mu\nu},g_{\mu\nu})+\xi^{\mu}(x,z)H_{\mu}(\pi^{\mu\nu},g_{\mu\nu}),$$
which is a sum of just constraints. These constraints are the dual versions of the anomalous Ward identity for broken Weyl invariance and for diffeomorphism invariance on the side of the quantum field theory.
\begin{equation}
H(\pi^{\mu\nu},g_{\mu\nu})=\textrm{tr}\pi-f(\pi^{\mu\nu},g_{\mu\nu}),
\end{equation}
\begin{equation}
H_{\mu}(\pi^{\mu\nu},g_{\mu\nu})= -2\nabla_{\rho}\pi^{\rho}_{\mu}.
\end{equation}
The question then is what form the function $f(\pi^{\mu\nu},g_{\mu\nu})$ can take, and what dictates it's structure. In the following subsection, I will show how the demand that the Poisson algebra of these constraints agrees with that of general relativity fixes completely the form of this function and when translated back to the field theory, this will correspond exactly to the desirable form of the anomalous Ward Identity.
\subsection{The constraint algebra}
Consider the smeared forms of the constraints
$$H(\sigma)=\int \textrm{d}^{2}x \sigma(x)H,$$
$$H_{\mu}(\xi^{\mu})=\int \textrm{d}^{2}x \xi^{\mu}(x)H_{\mu}.$$

If we aim at reconstructing general relativity in the bulk, then it will be of useful to note that the Poisson bracket algebra 

\begin{equation}\left\{H_{\mu}(\xi^{\mu}),H_{\mu}(\chi^{\mu})\right\}=H_{\mu}([\xi,\chi]^{\mu})\end{equation}
\begin{equation}\left\{H(\sigma),H_{\mu}(\xi^{\mu})\right\}=H(\xi^{\mu}\nabla_{\mu}\sigma)\end{equation}
\begin{equation}\left\{H(\sigma),H(\sigma')\right\}=H_{\mu}(g^{\mu\nu}(\sigma\partial_{\nu}\sigma'-\sigma'\partial_{\nu}\sigma)),\end{equation}
is restricting enough to fix the form of the total Hamiltonian, and consequently of both $H$ and $H_{\mu}$ to be of the form
\begin{equation}H=\frac{1}{\sqrt{g}}G^{\textrm{dW}}_{\mu\nu\rho\sigma}\pi^{\mu\nu}\pi^{\rho\sigma}-\sqrt{g}R+\textrm{tr}\pi\label{hc}\end{equation}
\begin{equation}H_{\mu}=-2\nabla_{\nu}\pi^{\nu}_{\mu}.\end{equation}
The Ward identity corresponding the the diffeomorphism invariance of the boundary field theory gives us the vector constraint, but the form of the anomalous Ward identity for broken Weyl invariance need be fixed by the constraint algebra. So in other words, the constraint algebra forces 
\begin{equation}f(\pi^{\mu\nu},g_{\mu\nu})=\frac{1}{\sqrt{g}}G^{\textrm{dW}}_{\mu\nu\rho\sigma}\pi^{\mu\nu}\pi^{\rho\sigma}-\sqrt{g}R.\end{equation}
If we translate back through the dictionary from the bulk momentum to the (VEV of the) boundary energy momentum tensor, then it is seen that this is consistent with the form of the exact RG equation \cite{Ver} which had in it two constants $c$ and $\mu$ which were absorbed into the definition of the momenta and the normalisation of the overall scalar constraint. This has the same effect as setting $c=\frac{24\pi}{\mu}$, \footnote{Also notice that with this choice, the large $c$ and small $\mu$ limits are consistent with each other.}  which appears then as the overall normalisation constant of the bulk action. This fact will be important in the next section. 
So the constraint algebra is powerful enough to fix the form of the constraints themselves. This result is due to Hojman, Kuchar and Teitelboim \cite{HKT}. In the case of the vector constraint which is the bulk manifestation of the diffeomorphism Ward identity, this is perhaps unsurprising. For this $D$ dimensional diffeomorphism invariance to be enhanced into the full $D+1$ dimensional diffeomorphism invariance of the emergent bulk spacetime is a key necessity for theories that possess holographic duals, and this requires the scalar Hamiltonian constraint to take the same form as that of GR in the Hamiltonian formalism, first discovered by Arnowitt, Deser and Misner in \cite{ADM}. This is the bulk manifestation of the anomalous Ward identity corresponding to broken Weyl invariance. This is the Ward identity that corresponds to the response of the quantum field theory to the local RG transformations themselves, and the scalar constraint of GR that fixes $f(g,\pi)$ to take a very particular form is thus a signature of theories possessing bulk gravity duals. The statement now is that the constraint algebra is in and of itself powerful enough to fix the form of this Ward identity completely. On the field theory side, this translates into a statement regarding the Wess--Zumino (WZ) consistency conditions:
\begin{equation}
0=[\Delta_{\sigma},\Delta_{\sigma'}]\textrm{ln}Z_{QFT}[g]|_{c\rightarrow \infty}=\int \textrm{d}^{2}x \sqrt{g} g^{\mu\nu}(\sigma \partial_{\nu}\sigma'-\sigma' \partial_{\nu}\sigma)\langle \nabla_{\kappa}T^{\kappa}_{\nu}\rangle.
\end{equation}
I will dub these the `holographic' Wess--Zumino consistency conditions \cite{me}.

In the situation of interest, this condition on the field theory when satisfied ensures the form of the exact RG equation that correspond to the Hamilton--Jacobi equation in the bulk gravitational theory.

\end{subsection}
\subsection{The Hamilton--Jacobi Equation}
Give the normalisation of the bulk action that has been chosen, we see that the role of $\hbar$ in the bulk theory is played by $\frac{1}{c}$. In the large $c$ limit, this is small and the functional integral can be evaluated in the saddle point approximation. In other words, the bulk theory is in its semiclassical limit. This implies that the relationship
\begin{equation}\lim_{c\rightarrow \infty}\textrm{ln}Z_{QFT}[g]\propto \frac{c}{24\pi}S_{B}^{o.s.},\label{ehj}\end{equation}
holds, where $o.s.$ superscript denotes that the bulk action is taken on shell. The Hamiltonian constraint then takes the form of the Einstein-Hamilton-Jacobi equation 
$$
g_{\mu\nu}\frac{\delta S^{o.s}_{B}}{\delta g_{\mu\nu}}=-R(g)-G^{\textrm{dW}}_{\mu\nu\rho\sigma}\frac{\delta S^{o.s}_{B}}{\delta g_{\mu\nu}}\frac{\delta S^{o.s}_{B}}{\delta g_{\rho\sigma}}.
$$
Which is simply the Hamiltonian constraint re-written with the identification $\pi^{\mu\nu}=\frac{\delta S^{o.s}_{B}}{\delta g_{\mu\nu}}$ being made. Given the identification \eqref{ehj}, this implies that:
$$\pi^{\mu\nu}=\lim_{c\rightarrow \infty}\frac{24\pi}{c}\frac{\delta }{\delta g_{\mu\nu}}\textrm{ln}Z_{QFT}[g]=\langle T^{\mu\nu}\rangle.$$

As promised, this is nothing but the factorised exact RG equation
\begin{equation}
\langle T^{\alpha}_{\alpha}\rangle=-R(g)-G^{\textrm{dW}}_{\mu\nu\rho\sigma}\langle T^{\mu\nu}\rangle\langle T^{\rho\sigma}\rangle,
\end{equation}
with no corrections. 

In the following subsection, I will describe where the cosmological constant lies in this construction. 
\subsection{The cosmological constant}
In order to bring the Hamiltonian constraint to a more standard form, it is necessary to eliminate the term linear in the momentum. This will require performing a canonical transformation:
\begin{equation}
\pi^{\mu\nu}\rightarrow \pi^{\mu\nu}-\frac{\delta C[g]}{\delta g_{\mu\nu}},
\end{equation}
with generating functional $C[g]=\frac{24\pi}{c} \int \textrm{d}^{2}x \sqrt{g}$ proportional to the volume. The price of performing this canonical transformation is that the momentum independent part of the scalar constraint is modified as follows:
\begin{equation}
\sqrt{g}R(g)\rightarrow \sqrt{g}\left(R(g)\right)-\frac{1}{2\sqrt{g}}G^{\textrm{d.W}}_{\mu\nu\alpha\beta}\frac{\delta C[g]}{\delta g_{\mu\nu}}\frac{\delta C[g]}{\delta g_{\alpha\beta}}=\sqrt{g}(R(g)-2).
\end{equation}
Thus we see that the net effect of the canonical transformation is to add a cosmological constant to the theory (in units where it has been set to 1). The overall normalisation of the action is thus consistent with the identification $c\propto \frac{l}{G}$.

Also note that the momenta $\pi^{\mu\nu}$ can now be integrated out in the bulk partition function, which in the saddle point approximation is nothing but the Legendre transform. We then find that  
\begin{equation}S_{B}=\frac{c}{24\pi}\int \textrm{d}^{3}x\ (^{(3)}R(\gamma)-2)\sqrt{\gamma}\label{eh} .\end{equation}
which is nothing but the Einstein--Hilbert action for the three dimensional metric $\gamma_{ab}$ on the bulk space with negative cosmological constant set to 1.

\section{Conclusion and Summary}
In this article, I presented the construction of the off shell Einstein--Hilbert action \eqref{eh} from the RG flow of a large $c$ $2D$ CFT deformed by the $T\bar{T}$ operator. This can also be seen as a different perspective on the $T\bar{T}$ deformation itself which has gained quite a bit of attention recently. This is one of very few examples of an irrelevant RG flow that seems to be under control in some sense. For instance, the
modification of the finite size energy spectrum for such theories can be computed \cite{Zamo1}, \cite{Ital}. 
From a holographic perspective, this deformation is also expected to modify the asymptotic behaviour of the bulk theory under consideration, see for instance \cite{LST}. The perspective I offer differs from this, because through the QRG mapping, this deformation is exactly what is needed to provide a kinetic term for the off shell ADM Hamiltonian constraint, which is an off-shell object.

 Assuming that the holographic Wess--Zumino consistency conditions are satisfied, the deformation under consideration remains un corrected by irrelevant operators of higher mass dimensions and gradients etc. And so, the coupling of the operator $\mathcal{O}$ still provides the only scale deforming the theory. What's more, the curved space generalised version of the factorisation property $$\langle \mathcal{O}\rangle=G^{\textrm{d.W}}_{\mu\nu\rho\sigma}\langle T^{\mu\nu}\rangle\langle T^{\rho\sigma}\rangle$$ is also a consequence of this condition. 
 In other words, the HWZ condition `picks out' the $T\bar{T}$ deformation, and the resulting bulk theory is GR in the ADM formulation. Information about any particular holographic theory provides initial conditions for a kind of initial value problem for the bulk gravity theory. \footnote{Given that the emergent direction considered here is Euclidean this isn't a Cauchy problem per se.} So instead of thinking of this deformation as modifying the asymptotics in the bulk theory, it can also be seen as providing initial conditions for the evolution problem which can be solved to describe the radial development of some given constant radius hypersurface. The question of asymptotic can only be answered if one understands the long time evolution properties of any particular solution. 
 
It is worth noting that the QRG kernel is well defined and understood in the large $c$ limit, but going beyond this, there is a kernel defined by Freidel in \cite{laurent} which ensures that any CFT partition function can be mapped into a pair of `radial' states in a bulk quantum gravity theory with negative cosmological constant. Making contact with this result will be an interesting topic for future research. 

\section*{Acknowledgements}
I thank Lee Smolin for comments on an early version of this manuscript and Jaume Gomis for engaging discussions regarding the $T\bar{T}$ flow. I also thank Mark Mezei for an enlightening email exchange. 

This research was supported in part by Perimeter Institute for Theoretical Physics as well as by grant from NSERC and the John Templeton Foundation. Research at
Perimeter Institute is supported by the Government of Canada through the Department of Innovation,
Science and Economic Development Canada and by the Province of Ontario through the Ministry of
Research, Innovation and Science

\end{document}